\documentclass[twocolumn]{emulateapj}
\def \ellipse {{\sc Ellipse}}
\def \stsdas {{\sc Stsdas}}
\def \sext {{\sc Sextractor}}
\def \HST{{\emph{HST}}}

\begin{document}

\slugcomment{12/06/06}

\title{Radius Dependent Luminosity Evolution of Blue Galaxies in GOODS-N}

\author{J. Melbourne, A. C. Phillips, J. Harker, G. Novak, D. C. Koo, \& S. M. Faber}
\affil{UCO/Lick Observatory, Department of Astronomy and Astrophysics, 
University of California at Santa Cruz, 1156 High St., Santa Cruz, CA 95064}
\email{jmel, phillips, jharker, novak, koo, faber@ucolick.org}

\begin{abstract}
We examine the radius-luminosity (R-L) relation for blue galaxies in the Team Keck Redshift Survey (TKRS) of GOODS-N.   We compare with a volume-limited, Sloan Digital Sky Survey sample and find that the R-L relation has evolved to lower surface brightness since $z=1$. Based on the detection limits of GOODS this can not be explained by incompleteness in low surface-brightness galaxies.  Number density arguments rule out a pure radius evolution. It can be explained by a radius dependent decline in $B$-band luminosity  with time.  Assuming a linear shift in $M_B$ with $z$, we use a maximum likelihood method to quantify the evolution.  Under these assumptions, large ($R_{1/2} > 5$ kpc), and intermediate sized ($ 3 < R_{1/2} < 5$ kpc) galaxies,  have experienced  $\Delta M_B =1.53 (-0.10,+0.13)$ and $1.65 (-0.18, +0.08)$ magnitudes of dimming since $z=1$. A simple exponential decline in star formation with an e-folding time of 3 Gyr can result in this amount of dimming.  Meanwhile, small galaxies, or some subset thereof, have experienced more evolution, $2.55 (\pm0.38)$ magnitudes.   This factor of ten decline in luminosity can be explained by sub-samples of starbursting dwarf systems that fade rapidly, coupled with a decline in burst strength or frequency.  Samples of bursting, luminous, blue, compact galaxies at intermediate redshifts have been identified by various previous studies.  If there has been some growth in galaxy size with time, these measurements are upper limits on luminosity fading.

\end{abstract}

\section{Introduction}
Hierarchical clustering models (e.g. Baugh et al. 1998;  Somerville, Primack \& Faber 2001) predict that galaxies assemble by the merger of smaller components with much of the size evolution of galaxies occurring since redshift $\sim3$.  Merger histories of galaxies can be traced by observables such as half-light radius, stellar luminosity, stellar mass and metal abundance.  In this paper 
we look for evidence of size dependent, $B$-band luminosity evolution within blue galaxies to $z=1$.

The study of distant galaxy sizes first became possible in the 1990's with the advent of high resolution Hubble Space Telescope (\HST) imaging.   Lilly et al. (1998) used WFPC2 F814W images to investigate the radius-luminosity (R-L) relation for 341 galaxies in the Canada France Redshift Survey (CFRS).   They found that large spiral galaxies undergo  little size evolution between $z=1$ and today.  However they found evidence for $\sim1$ magnitude of surface brightness evolution (at fixed radius) for those same systems.  They also indicated that smaller systems may be experiencing more evolution than large ones, but they did not quantify this statement.  Simard et al. (1999) performed a similar study in a magnitude limited sample of 190 galaxies in the Groth Survey Strip ($I_{814} < 23.5$).  While they found evidence for roughly 1.3 magnitudes of surface brightness evolution (to $z=1$), they claimed that most of the apparent evolution could be produced by incompleteness effects.  This conclusion was questioned by Bouwens and Silk (2002) in a paper presenting results from simulations of galaxy formation.  They predicted 1.5 magnitudes of evolution in the surface brightness of galaxies from $z=1$ to today.  They suggested that observed apparent evolution in the Simard et al. data set may be real and not the result of selection biases.

With the refurbishment of \HST\ and the installation of the Advanced Camera for Surveys (ACS), such studies are now possible on a larger scale. Ravindranath et al. (2004), in an analysis similar to Simard et al. (1999), reported on the evolution of the radius-luminosity relation for disk galaxies in the Great Observatories Origins Deep Survey (Giavalisco et al. 2004) south field (GOODS-S).  Using photometric redshifts, they reached fainter apparent magnitudes than previous studies. However, they limited their analysis to a  small sub-sample of galaxies which they believed to be 90\% complete, based on the detection limits for smooth model galaxies.  Constructing luminosity functions, they found no evidence for evolution in $L^*$ to $z=1$. 

Barden et al. (2005), who studied the R-L relationship for 5664 disk galaxies in the GEMS field, however, argued that both Simard et al. and Ravindranath et al. results were biased by their strict surface-brightness cuts.  Specifically, both Ravindranath et al. and Simard et al. culled their data such that they were only studying the highest surface brightness tails of the  galaxy luminosity distributions at all redshifts.  Barden found that when they imposed the Ravindranath  et al. strict surface brightness cut on their sample, it eliminated $\sim70\%$ of the bright ($M_B <-20$) galaxies in the low-$z$ bin but only 10\%  of galaxies at high-$z$.  Because they were only looking at the tails of distributions there was no characteristic luminosity to compare too, resulting in a measurement of no surface brightness evolution.  However, when Barden et al. instead applied their own selection function to all galaxies brighter than their high-$z$ magnitude limit of $M_B <-20$, they found that the central surface brightness has evolved significantly, 1.4 magnitudes in the $B$ band out to $z=1$.    Barden et al. noted that their results mapped smoothly onto the R-L relation for nearby Sloan Digital Sky Survey (SDSS) galaxies, which was not true for the Ravindranath et al. and Simard et al. results.  Interestingly, Barden et al. also showed evidence for an unchanging stellar mass-size relation for disk galaxies since $z=1$.  However, because galaxies are known to be star forming, the stellar mass must be increasing, indicating that the size of galaxies must also be increasing to remain on the mass-size relation.  They suggest that this is evidence for "inside-out" disk growth. 
 
A study by Trujillo \& Aguerri (2004) attempted to measure the size evolution of disk galaxies in the Hubble Deep Fields (Williams et al. 1996).  They compared the SDSS R-L relation from Shen et al. (2003) with 218 galaxies in the HDF and specifically looked at how the mean and dispersion of galaxy sizes in a given luminosity bin changed with redshift.  They found a modest increase in size with  time, which could also be interpreted as a decrease in $V$-band luminosity of 0.8 magnitudes since $z=0.7$.  Given a modest color evolution, this luminosity evolution was roughly in line with the Barden et al. results.  Trujillo \& Pohlen (2005) built upon the Barden et al. and Trujillo \& Aguerri results, measuring truncation radii and magnitudes of 36 galaxies in the \HST\ ultra-deep field (UDF).  After applying the Barden et al. luminosity evolution to the R-L distribution of their high $z$ sample ($z\sim1$) they claimed that an additional truncation radius growth of 25\% was necessary to match to a set of 35 local disk galaxies measured in the same way.  They suggested that this is further evidence for moderate inside-out disk growth.  
 
In this paper we attempt to measure the surface brightness evolution of blue galaxies in GOODS-N.  We use a different approach from these previous works.  First, we quantify incompleteness using real galaxies, rather than the smooth model galaxies used in previous studies.   For instance we compare the detection of galaxies in the UDF to galaxies in the same field at the GOODS depth and find that to our magnitude limit, GOODS is not missing a significant number of low surface-brightness galaxies.  Second, rather than apply a single set of selection criteria to all of the galaxies in our sample, we start with a statistically complete, \emph{volume-limited} local sample from SDSS that reaches very low luminosity.  We then ask by how much does the local sample need to evolve to match distributions drawn from GOODS at successively higher redshift. 
 
We use the SDSS to model the shape and number distribution of the local R-L relation and to anchor our evolution measurements. By drawing on a statistically complete, \emph{volume-limited}, sample of galaxies reaching to very faint luminosity levels ($M_B < -16$), we can identify specific features in the luminosity distributions of galaxies that can be used to track evolutionary trends.  Two such features are seen in SDSS, 1) the peak of the magnitude distribution for galaxies with large half-light sizes ($>5$ kpc); and 2)  the sharp decline in the numbers of more luminous galaxies that have smaller radii.  

We compared the local SDSS R-L relation to 931 blue galaxies from the GOODS-N field with spectroscopic redshifts from Team Keck Redshift Survey (TKRS, Wirth et al. 2004). We measured elliptical aperture photometry of the GOODS galaxies from the very deep \HST\ imaging obtained with ACS.  The four ACS bands spanning observed blue to near-IR provide excellent  constraints on photometric K-corrections, which are significant at higher redshifts.  The depth of the GOODS imaging allows for a very accurate estimate of the incompleteness biases in the sample, which is vital for determining the validity of any apparent evolution.  Spectroscopic redshifts from TKRS, obtained with the DEIMOS spectrograph, are a significant improvement on photometric redshifts for very blue systems, for which photo-$z$'s are difficult to measure.  Because the  TKRS sample is large enough to divide into different categories, we can, for example,  look for luminosity evolution as a function of galaxy size, or galaxy size evolution as a function of luminosity. Neither of these two programs have been well  explored in the past. The size and depth of the SDSS and TKRS surveys gives us the current best data sets for tracing the evolution of galaxy luminosity and size. 

Data for both the local and high redshift galaxy samples and discussion of selection effects are presented in \S\,2. Radius-luminosity distributions and evolution of each parameter are detailed in \S\,3.  Discussion of the implications of the observed evolution is provided in \S\,4. Apparent magnitudes are reported in the AB system, while absolute magnitudes are given on the Vega system.  We assume a flat cosmology with $H_0= 70$ km/s/Mpc, $\Omega_m=0.3$, and $\Omega_{\lambda}=0.7$.

\section{The Data}
\subsection{GOODS Imaging and TKRS Spectra}
The GOODS team obtained very deep imaging data in two $10\arcmin \times 16\arcmin$ fields using the \HST\  Advanced Camera for Surveys. This effort included four separate passbands; F435W, F696W, F775W, and F850LP ($B,V,i$ and $z$), which allow for good matching to spectral energy distributions, and determination of K-corrections.    Our study uses the mosaic data ($0.03 \arcsec$ pixels) from the northern field (GOODS-N), which is centered  on the Hubble Deep Field North but covers a much larger area. 
   
We limit our study to those GOODS galaxies targeted by  TKRS (Wirth et al. 2004), which obtained spectra of 2018 objects in the GOODS-N field.  This magnitude-limited survey targeted objects with $R_{AB} \le 24.4$, identified on deep ground based images.  TKRS made no effort to exclude objects that appeared star-like, so the survey is not biased against compact galaxies.  The survey obtained reliable redshifts for 1440 galaxies and spectroscopically identified 96 Galactic stars.

\subsection{Photometry}

\subsection{Elliptical Aperture Photometry}
We measured photometric properties of the TKRS sample with the \stsdas\ program \ellipse\, which provides flux and intensity measurements within elliptical apertures of increasing semi-major axis. \ellipse\ performed between 1000 to 2000 iterations to fit the central coordinates, ellipticity and position angle at every semi-major axis radius. Radii were incremented with a geometrical step of 0.0015\arcsec from a minimum radius of 0.003\arcsec to a maximum of 6\arcsec, well out to the sky for all but the largest low-$z$ elliptical galaxies in our sample. The centers and ellipticities were allowed to vary at each radius.  We did the initial fitting with the $i$-band image, the deepest of the four bands.  The $i$-band ellipses were then applied for photometry in the other three bands.

While running \ellipse, neighbor galaxies were masked out with  \sext\ (Bertin \& Arnouts 1996) segmentation images.  We set the \sext\ detection parameters to maximize the masking of neighbors, while retaining the structure within a given galaxy.  Specifically, we set the {\sc DETECTION} and {\sc ANALYSIS THRESHOLDS} to 2.3 sigma, and the {\sc DEBLEND\_MINCONT} to 0.10.  The {\sc DETECTION\_MINAREA} was set to 25 pixels.  Final segmentation maps were created by adding together the maps from the $V$ and $i$-band images.  We found that while the $i$-band map did a better job of detecting the wings of galaxies, the $V$-band image was important for identifying regions of young stars.  After combining the segmentation maps from the two filters, we smoothed them with a boxcar smooth of 10 pixels.  This, for instance, filled in the pixels between spiral arms that might not have been included in the original maps and extend the segmentation images into the lower surface brightness wings of the galaxy.  Final maps were inspected by eye and any pixels that were masked but belonged to the actual galaxy being measured were corrected.
      
\subsection{Magnitudes and Half-Light Radii}
Total magnitudes, half-light radii, concentration parameters, and surface brightnesses were calculated from the \ellipse\ measurements using a curve of growth technique. For each image, we employed an iterative process.  An initial sky level was chosen as the flux level at which the intensity of light in successive annuli remained constant.  After subtracting the sky value, a total galaxy flux was determined as the flux level at which the enclosed flux remained constant in successive apertures.  If the enclosed flux was found to fall off with large radius, the sky level was reduced and a new total flux calculated.  If the enclosed flux was found to increase with radius without bound, the sky level was increased and the process begun again.

Neighbors were often near enough to contaminate the sky background measurement.  \sext\ segmentation images were used to identify and mask out these contaminants.  Cases where the masking failed to subtract the entire influence of a neighbor were identified by inspection and corrected individually.

We converted apparent magnitudes to absolute $M_B$ [Vega] and rest-frame colors, $U-B$ and  $B-V$ [Vega], using the  K-correction routine described in Willmer et al. (2005).  

Apparent half-light radii ($r_{50}$) were given by the semi-major axis that enclosed half the flux. We corrected apparent radii for the point-spread-function of the image by subtracting the radii of stellar sources in quadrature (Phillips et al. 1997).  To avoid possible variations of radius with passband, we measured size in all four filters and then estimated the  rest-frame $B$-band half-light radius from a weighted mean of the results.  Weights were assigned by the overlap of the filter passband with the rest-frame $B$ filter.  Using the angular-diameter distance, we calculated rest-frame $B$-band half-light radii in kpc ($R_{1/2}$).   A concentration measurement was made from the $i$-band images, $C_{80/20}=\mbox{log}(r_{80}/r_{20})$, where $r_{80}$ is the radius that contains 80\% of the galaxy light. 

Table \ref{table:ktrs} summarizes the results of the elliptical photometry of the TKRS sample of galaxies.  Column 1 is the TKRS ID number (Wirth et al. 2004); columns 2 and 3 contain the R.A. and Dec, respectively; column 4 gives the redshift (Wirth et al. 2004); and columns 5 - 8 give the \ellipse-measured magnitudes in the 4 GOODS bands.  Columns 9 and 10 are $M_B$ and $(B-V)_{rest}$, respectively;  column 11  contains the $i$ band apparent half-light radius, $r_{50}$ [$\arcsec$]; and column 12 gives the rest-frame $B$-band half-light radius, $R_{1/2}$ [kpc]. The full table is available in the online edition of The Astrophysical Journal.

\begin{figure}
\includegraphics[trim=90 100 0 150, clip, scale=0.6]{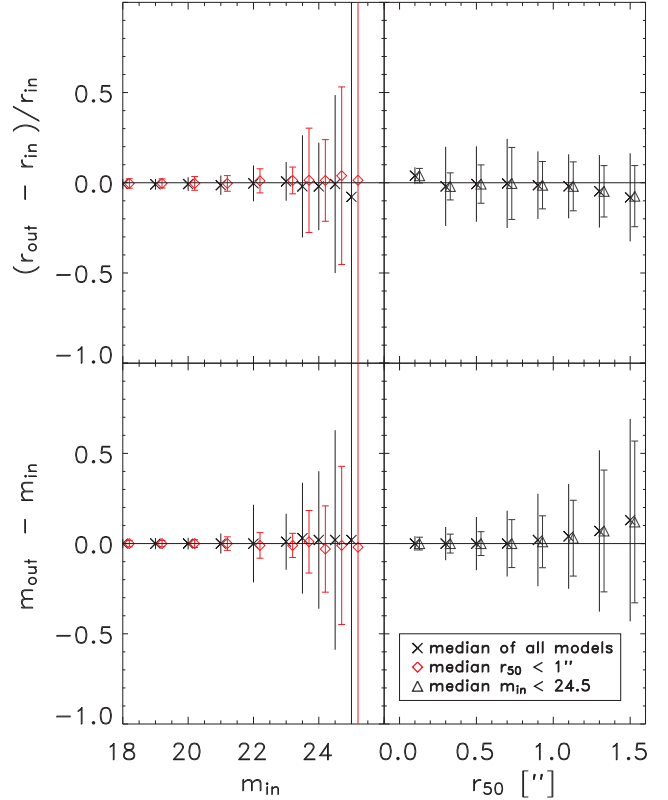}
\figcaption{\label{fig:model_phot} A comparison of "measured"  vs. "true" galaxy parameters for 3000 model exponential disk galaxies placed into the GOODS $i$-band images and measured with the IRAF \ellipse\ fitting routine.  Fractional errors in the parameters are plotted against input magnitude (left panel) and half-light radius (right panel).  The top panel shows apparent half-light radius, and the bottom shows apparent magnitude.  Ten model galaxies were generated for every input size, brightness, and ellipticity and the models span the range of parameters exhibited in the TRKS galaxy sample.   Black x's represent the median of all 300 models within a bin.   Red diamonds represent models with input half-light radii smaller than $1\arcsec$, while blue triangles are models brighter than 24.5 mags. RMS errors are shown as vertical lines.  ELLIPSE does a good job measuring galaxy magnitudes except at the faintest end.  }
\end{figure}

\subsection{Selecting Blue Galaxies}
Recent studies (Blanton et al. 2003A; Bell et al. 2004; Weiner et al. 2005) demonstrate that galaxy populations are bimodal in optical color.   The color-magnitude diagram for TKRS also shows that galaxies fall into two regions in color space, blue galaxies and red sequence galaxies, with a trough separating the two groups.  The color of this trough is roughly $(B-V)_{rest} = 0.7$ [Vega] for the TKRS sample.  Locally in the trough is redder by about 0.1 mags.  Red galaxies tend to be ellipticals  and early-type, bulge-dominated spirals, both with high central surface brightness.  The TKRS red sequence galaxies, for instance, tend to have high concentrations, $C_{80/20}\sim 0.9$, compared with the typical blue galaxy, $C_{80/20}\sim 0.6$.  Therefore the red galaxies tend to follow a separate R-L relation from blue galaxies.   As this would complicate our analysis, we choose to leave the red galaxies out of our sample. Except where noted, the rest of this paper discusses the TKRS galaxies bluer than the trough ($(B-V)_{rest} < 0.7$), which tend to be later-type spirals, irregulars, and compact blue galaxies.  We also eliminate ten obvious AGN galaxies based on their images and spectra. 

\subsection{Photometric Accuracy}


We tested for the photometric accuracy of our measurement methods in two ways.  First we created idealized exponential disk model galaxies with known sizes and luminosities and measured them in the same way as the GOODS galaxies.  However, because real blue galaxies often contain structure from star forming regions, spiral arms or bars, our "smooth" model galaxies may not be adequate for understanding our measurement errors.  As a result we also performed an additional test using real galaxies in the Ultra Deep Field (UDF; GO 9978; principal investigator S. Beckwith).  In this second test we compared measurements of galaxies observed in the UDF (two magnitudes deeper than GOODS) with measurements of the same galaxies observed to the GOODS depth.  If our techniques are miss-measuring the low surface-brightness components of galaxies there should be an offset in the sizes and luminosities of galaxies measured at the two significantly different depths.  While the measured radii of our large low-surface brightness model galaxies tend be underestimated, the radii of actual galaxies do not seem to have this bias.  The details of these studies are discussed in the following two subsections.    

\subsubsection{Photometry of Model Galaxies}
To check the accuracy of our photometry, we measured  over 3000 model disk galaxies, spanning the range of apparent magnitudes ($i=18-25$), radii ($r=0.1-1.5 \arcsec$), and ellipticities ($e=0.3-0.9$) exhibited in the TKRS galaxy sample.  For simplicity, single-component exponential disks were generated using the IRAF MKOBJECT routine.  Galaxies were convolved with a gaussian PSF with $0.06 \arcsec$ radius.  Models were placed at random in the GOODS $i$-band images, however models that were obviously placed on top of existing galaxies were not measured.  For every input magnitude, radius, and ellipticity, 10 galaxies were generated.  A postage stamp of each model galaxy was made and measured in the same way as the TKRS galaxies.  A summary of the resulting photometry of model galaxies is given in Figure \ref{fig:model_phot}.

The left panel compares the input vs. output photometry as a function of apparent magnitude, while the right panel plots the same as a function of input radius.  
Output magnitudes, especially for galaxies smaller than $1\arcsec$, match well to the input mags (better than 0.1 mags for $m_{in} < 23$) with increasing scatter towards the faint end.  Sizes are also well reproduced ($<10\%$).   Again the scatter increases for faint objects.  
Looking at the accuracy of the photometry as a function of input galaxy radius (right panel), 
we find that the largest objects tend to have under-estimated radii and luminosities, because the low surface brightness disks tend to blend into the sky background.

The results of the model photometry indicate that for most isolated sources, our methods produce accurate measurements (luminosity to better than 0.1 mag and radii to better than 10\% for $m_{in} < 23$).  As the input radii and magnitudes increase, the scatter rises significantly. However, by taking groups of galaxies in aggregate, we can expect a representative picture of the galaxy population above our magnitude cutoff out to $z=1$.  While these model galaxies were made with a Gaussian PSF, we checked to make sure that our measurements would not change significantly for ACS PSF's which can have significant wings.  We created 10 additional models spanning our range of size of luminosity and convolved with the ACS $z$-band PSF from Tiny-Tim (Krist 1995).  We measured them in the same way as the original models and found no significant difference in the photometry.

\begin{figure}
\includegraphics[trim= 0 0 0 0, scale=0.6]{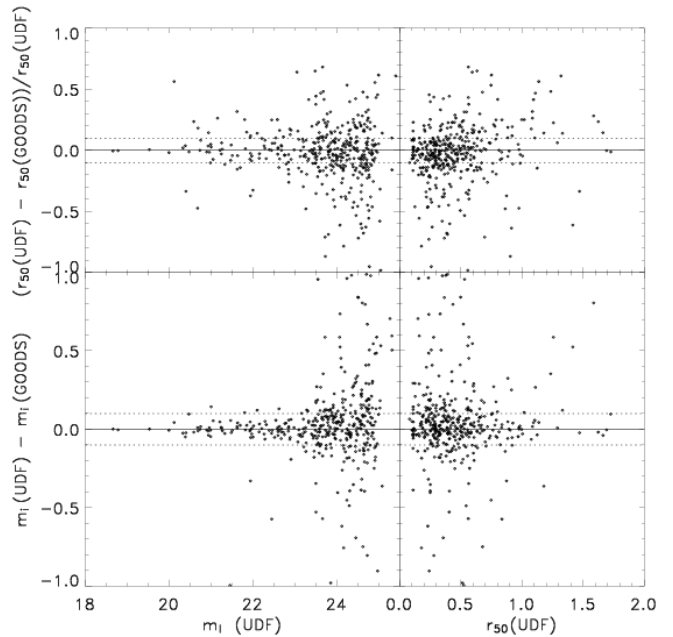}
\figcaption{\label{fig:udf} This plot is similar to Figure \ref{fig:model_phot} only now we are comparing the measured photometry of galaxies in the UDF to the same galaxies observed at the depth of the GOODS survey.  We show the comparison of both sizes and luminosities as a function of size and luminosity.  The Figure shows that there is not an offset between the GOODS and UDF photometry, and for $i < 23$ they agree to within 10\%.  Thus the GOODS photometry does not suffer any bias from low surface-brightness components to the galaxies.
}
\end{figure}

\subsubsection{Photometry in the UDF}
In the previous section we tested our photometric methods on smooth model galaxies of known size and luminosity.  While this test would have revealed any significant measurement errors for bright systems, it may not tell us much about measuring actual galaxies especially at fainter luminosities.   Actual galaxies can have disturbed morphologies and light profiles that are not well fit by smooth profiles.  We looked for additional measurement biases in our photometry by measuring galaxies in the UDF.  Galaxies with an apparent luminosity $i < 25$, were identified in the UDF with the \sext\ search algorithm.  We measured the apparent size and luminosity of these galaxies in both the UDF and the GOODS depth image of the UDF.  Because the UDF is two magnitudes deeper than the GOODS equivalent image, low surface brightness features will be easier to measure in the UDF.  

Figure \ref{fig:udf} shows a comparison of photometry from the UDF with the GOODS equivalent of the UDF.  For bright systems, $i < 23$, measurements of magnitude and size are equivalent to within 10\%.  For fainter systems the scatter increases, but is not biased.  This is another indication that low surface brightness wings of galaxies are not being missed by the GOODS images, and our photometry is accurate even to faint levels. 

\begin{figure}
\includegraphics[trim=40 150 0 100,scale=0.55]{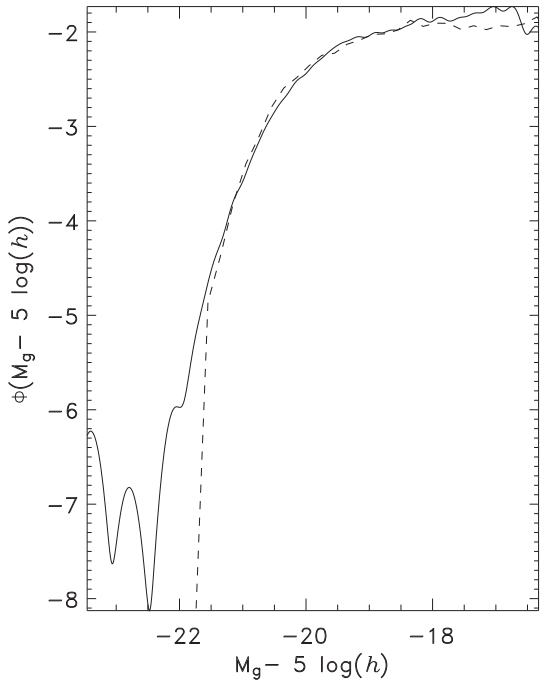}
\figcaption{\label{fig:lf}  To check whether our \emph{volume-limited} SDSS sample is representative of the local universe, we create a luminosity function from it and compare to Blanton et al. (2003b).  This figure plots ln($\Phi$) $[\mbox{number}/\mbox{h}^{-3}/\mbox{Mpc}/\mbox{mag}]$ against $M_g - 5 \mbox{log h}$, for all (red and blue) SDSS galaxies in our sample, with $h$ normalized to 1.  The solid line is the result from Blanton et al.  The dashed line is our own measurement.  Except at the extreme bright end, where even SDSS suffers from small numbers, the two distributions match well.
}
\end{figure}

\subsection{Local SDSS Data}
In Section 3 we will use a complete, volume-limited SDSS galaxy sample as a local reference for evolution in the R-L plane. The steps to generate the local SDSS R-L relation are described below.  

We select a statistically complete sample of blue galaxies from the SDSS DR2 catalogue with redshifts $0.01 < z < 0.1$  and apparent Petrosian magnitudes, $14 <= r_P <= 18.0$.   Half-light radii and luminosities were measured by Simard et al. (2006 in prep) with the GIM2D (Simard et al. 1998) Sersic model fitting routine.  The semi-major axis, half-light radii were corrected for PSF effects.   Magnitudes were converted from Sloan $g$ and $r$ to $B$ and $V$ using Eqn. 23 of Fukugita et al. (1996).  

From this set we drew a volume-limited sample that accounted for both the low and high apparent magnitude selection criteria. To do this, we selected galaxies of a given absolute magnitude in the redshift slice for which they are complete.  For instance, we drew high luminosity galaxies from the high-$z$ end of our SDSS sample (i.e. $z\sim0.1$) and drew low luminosity galaxies from the low-$z$ end.   We scaled the numbers of galaxies in each narrow $z$-bin to a common volume.  In order to confirm that our \emph{volume-limited} SDSS number distributions are valid,  we generated a luminosity function (using both red and blue galaxies) and compared with Blanton et al. (2003b).   We convert our $B$-band luminosities to SDSS $g$-band by assuming a typical galaxy color for blue galaxies of $g-B = -0.35$ (Fukugita et al. 1995, Table 3).  Figure \ref{fig:lf} shows the Blanton et al. result (line) and ours (dashed).  The two luminosity functions are well matched except at the brightest end, where the large statistical variations are due to small numbers of sources.  This test confirms that our \emph{volume-limited} SDSS sample is indeed representative of the local universe.

\section{Selection Function}
\subsection{Detectability of Low Surface-Brightness Galaxies}
We attempted to quantify the incompleteness effects that might bias the detection of galaxies in R-L space.  Low surface brightness galaxies at high redshift are the hardest for \sext\ algorithms to detect and the hardest for spectroscopic redshift determinations.  As detailed below, we studied selection biases introduced by low-surface brightness galaxies in three ways and found that our sample was not missing large numbers of these galaxies.        

\begin{figure}
\includegraphics[scale=0.4]{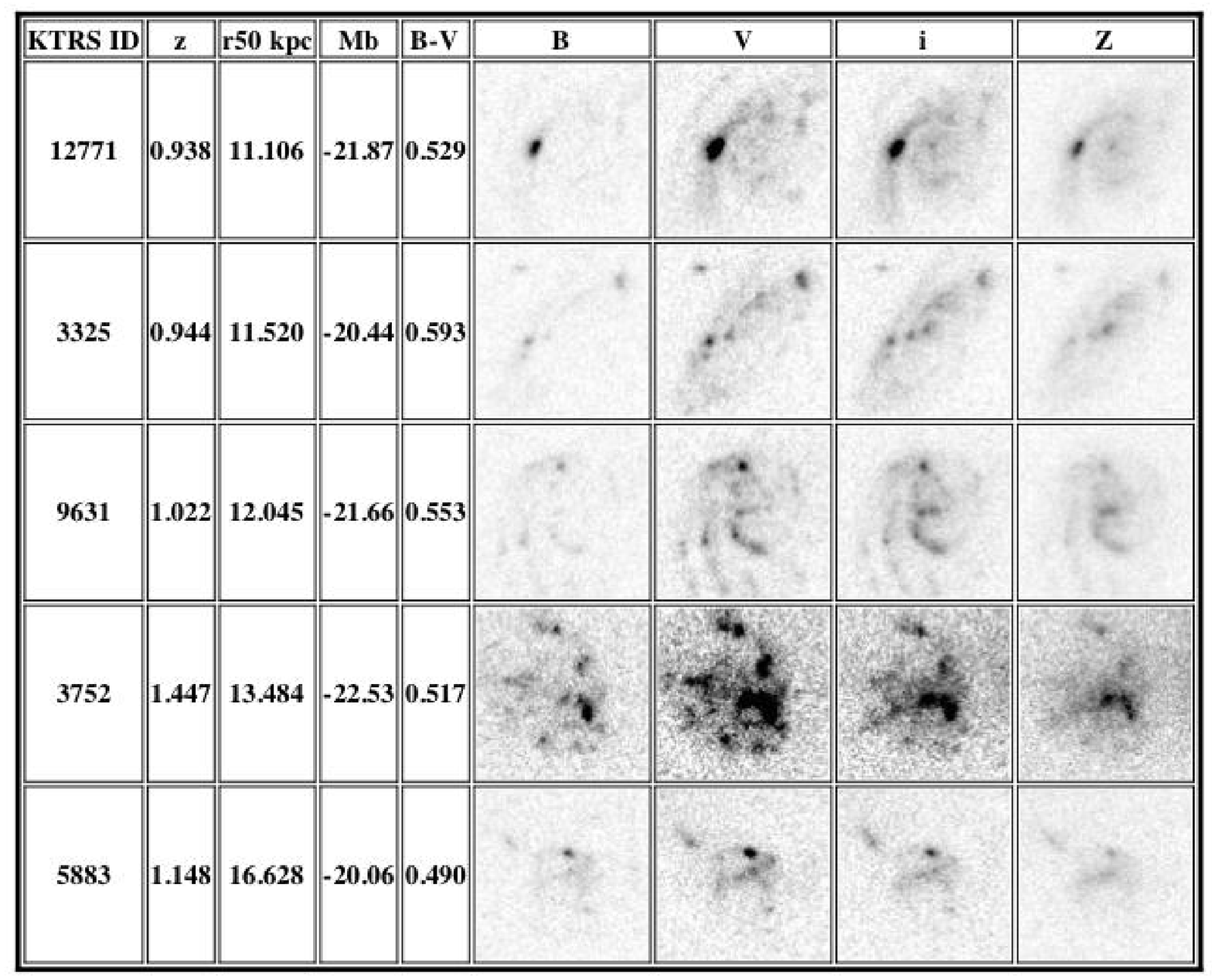}
\figcaption{\label{fig:bigones} Images of the five largest $z\sim1$ galaxies in GOODS-N.  These galaxies appear to be composed of multiple bright knots on top of disturbed, disky structures.  In general, they do not look like the normal disk galaxies of today. Our \ellipse\ measurements of their half-light radii, magnitudes, and colors are given.  The images are 3 arcsec on a side, which corresponds to $\sim24$ kpc at $z=1$.  Because they are composed of several high-surface brightness components, these galaxies are easy to detect.  If these objects are typical of large disk galaxies at $z=1$, then we are not missing a significant number of them.  Note: these images are somewhat smaller than the outer isophotes of the galaxies as measured by our process.
}
\end{figure}

First we investigated the detectability of the largest observed disk galaxies in the TKRS sample with redshifts near $z=1$. Figure \ref{fig:bigones} shows the five largest galaxies in the sample near $z\sim 1$.  These objects are not the relatively smooth disks that one finds locally.  In part because of the "morphological K-correction", they contain significant, high-surface-brightness substructure, which makes them much easier to detect than smooth disks.  In order to check if substructure can assist search algorithms, we extracted these galaxy images, doubled their diameter while conserving their flux, and  placed them back into the GOODS images at random locations.   All of these galaxies were still detected both visually and by \sext\ search algorithms. Furthermore, the multiple, compact, blue regions seen in these objects make them candidates for strong nebular emission lines, which allow for easy redshift determination.  If the five galaxies in Figure \ref{fig:bigones} are representative of large blue galaxies at $z=1$, then large galaxies are not being systematically missed in the GOODS catalogues.  In this case, several of the previous studies including Ravindranath et al. (2004) may have overestimated their selection effects at high-$z$ because they selected galaxy samples with cuts based on the detection of smooth, model galaxies rather than the knotty  galaxies that actually exist at high redshift. 

Second, we searched for low surface brightness galaxies in the GOODS fields that were not part of TKRS. We rebinned and smoothed the GOODS images and looked for objects that were not obvious in the original frames.  When we performed this experiment over $\sim1/3$ of the field with different smoothing lengths (up to $1\arcsec$), we found one additional large galaxy.  This galaxy was actually detected by the GOODS \sext\ catalogue, but as several objects rather than as a single galaxy.  Inspection by eye revealed that this object is most likely a single galaxy with multiple faint knots of star formation.

In a final check for low surface brightness galaxies, we ran the \sext\ search algorithm on both the UDF and the shallower GOODS images of the UDF.  Because the UDF reaches 2 magnitudes deeper than GOODS, any significant population of undetected low surface brightness galaxies should appear in these images.  Using the $i$-band images for detection, with an $i=25$ detection threshold, we found one additional low surface brightness galaxy in the UDF, that was not detected in the GOODS images.  This galaxy had an apparent magnitude of $i=24.8$, well below the TKRS magnitude limit.  Therefore, there is no evidence that GOODS is missing a significant population of low-surface galaxies to the magnitude limit of our sample.   

\subsection{The TKRS Selection Function}   
Redshift surveys are almost always biased against faint objects due to the increased difficulty of identifying spectroscopic features. To quantify selection biases in the TKRS catalogue, we compared the distribution of apparent radii and magnitudes for objects in GOODS-N with those in the TKRS catalogue.  We have already demonstrated that the GOODS \sext\ catalogues have no significant bias against low surface brightness galaxies down to the magnitude limit of TKRS.  Thus all that remains to check is what biases exist in the spectroscopic sample.  To do this we measured apparent size and luminosity for the $i < 25$ galaxies in the GOODS-N \sext\ catalogue.  The GOODS-N galaxies were measured in the same way as the TKRS sample, first running ELLIPSE and then our curve of growth code of postage stamp images of each galaxy.  

\begin{figure}[h]
\includegraphics[ scale=0.90]{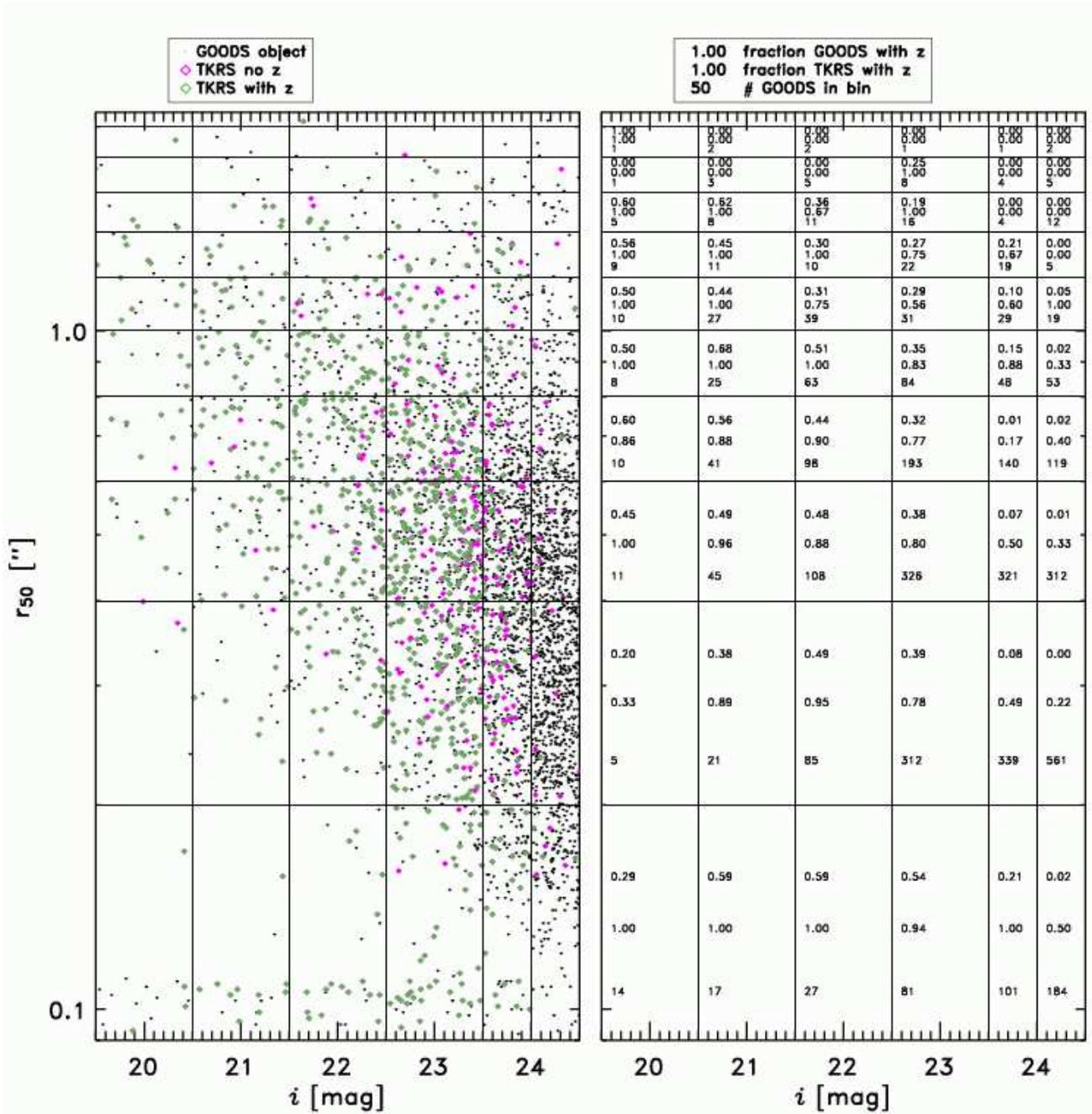}  
\figcaption{\label{fig:rl_ap} Apparent half-light radius is plotted against apparent magnitude for blue objects in the GOODS-N \sext\ catalogue (dots; Giavalisco et al. 2004). Objects for which TKRS obtained redshifts are shown in green.  Those that were targeted by TKRS without success are shown in red.  The figure is binned in both directions, and in the right hand plot, the ratio of successful $z$'s to GOODS objects is the top number in the bin, the ratio of of successful $z$'s to TKRS targeted galaxies is the middle number, and the total number of GOODS objects in each bin is the bottom number. Under the assumption that all the objects within a bin are in the redshift range of TKRS, the top number gives the TKRS selection function, and is what we use in our analysis. The major selection bias is against faint galaxies.  There is not a major trend with galaxy size.
}  
\end{figure}

The left panel of Figure \ref{fig:rl_ap} plots $r_{50}$ vs. $m_i$ for the GOODS-N galaxies.  Galaxies with successful TKRS redshift measurements are shown in green.  Those targeted but without a $z$ measurement are shown in red.  The plot is binned in both the radius and magnitude directions.  In the right panel of Figure \ref{fig:rl_ap} the ratio of successful $z$'s to GOODS galaxies in a bin is the top number in that bin.  The ratio of successful to targeted is the middle number, and the total number of GOODS galaxies in each bin is the bottom number.  

The plot shows that the major incompleteness occurs near the magnitude limit.  At a fixed magnitude, there is almost no drop off in success rate with radius, and thus with surface brightness.   Assuming that all of the galaxies in a bin are within the redshift range of TKRS, the top number in the bin summarizes the total TKRS selection function for galaxies of that size and brightness.  When measuring evolution in the R-L plane, we use the top number to correct for selection biases in our sample.  Because some of the galaxies without $z$'s may be at higher redshift than $z=1$, we may be over estimating the incompleteness correction. 

\section{Radius-Luminosity Relation}
\begin{figure}[h]
\includegraphics[scale=0.75]{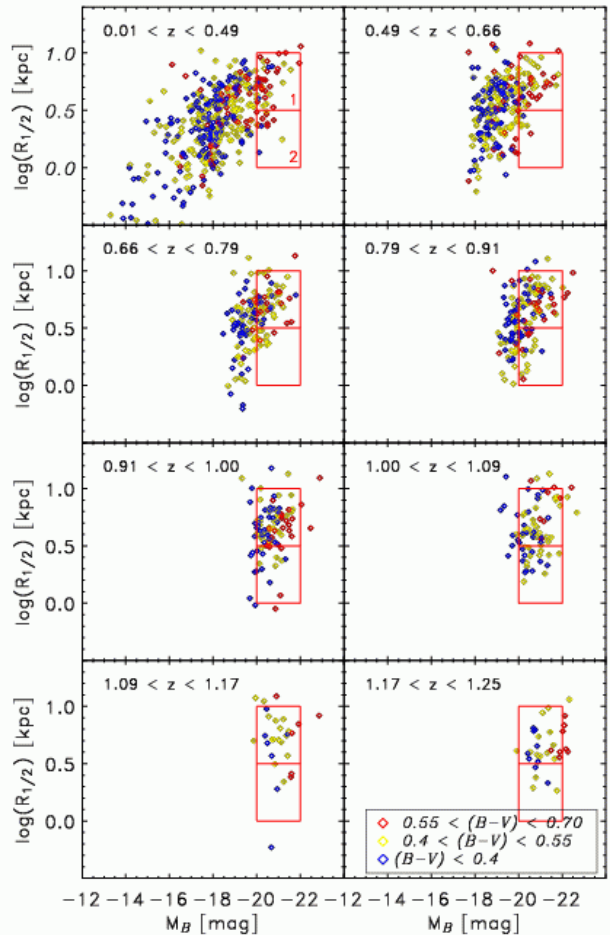}  
\figcaption{\label{fig:RL} Half-light radius of blue TKRS galaxies plotted against absolute magnitude in equal volume bins between $z=0$ and 1.2.   Red points have $(B-V)_{rest}$ color between 0.55 and 0.7.  Yellow points have $(B-V)_{rest}$ color between 0.4 and 0.55.  Blue points have color bluer than 0.4.  The red boxes are provided for reference.  If galaxies are fading with time they would tend to shift to the left in this plot, emptying boxes 1 and 2.  If galaxies are growing in size (but not in luminosity) they would tend to shift up in this plot, emptying box 2 before box 1.  Mergers would presumably move galaxies up and to the right over time.  In general, galaxies of a given radius in this ensemble appear to have faded since $z=1.25$ and galaxies in boxes 1 and 2 appear to have gotten redder.
}
\end{figure}

Figure \ref{fig:RL} shows the rest-frame radius-luminosity relation for blue galaxies  in TKRS to $z=1.2$. The redshift range of each bin has been selected to maintain \emph{equal volume}, $\sim3 \times 10^4$ Mpc$^3$.  The color of each galaxy is  indicated  and red boxes, labeled `1' and `2', are repeated in every panel to help guide discussion. The general trend, best seen in the lowest redshift bin, is that large galaxies tend to be bright and small galaxies tend to be faint.  In successive redshift bins, the sample has a low luminosity cut-off set by the TKRS apparent magnitude limit.  

Because the bins are equal volume, it is possible to visually compare the distributions and look for evolutionary trends, under the assumption that incompleteness effects are negligible. For instance, if there were no evolution in the R-L relation, we would expect that the number and distribution of objects in boxes 1 and 2 to  remain constant.  However, over time (from $z=1$ to the present), boxes 1 and 2 contain  successively fewer galaxies, suggesting evolution.  This is not the result of selection biases which would tend to produce the 
\clearpage
\clearpage
opposite effect. The evolution may be be occurring in one or more of the following parameters, luminosity, radius, or number density.    If we attribute the entirety of the evolution to a change in luminosity, galaxies are moving to lower luminosities (a shift to the left) over time, emptying the boxes.  Note that under the assumption that sizes are not changing, this is equivalent to the surface brightness evolution discussed in the literature.

In the case of pure radius evolution, galaxies would be growing hierarchically in size (a shift up in Figure \ref{fig:RL}) over time, emptying box 2 before 1.  A  model of pure radius evolution is somewhat unphysical in that the growth of galaxy stellar mass at large radii would presumably also increase luminosity.  In order for galaxies to change only in size and not in magnitude, the increase in luminosity from new star formation would  have to be balanced by a simultaneous fading of the other stars in the galaxy. While contrived, it will be interesting to investigate if the data can rule out a pure radius evolution scenario.  

The third case, number-density evolution would reduce the numbers of galaxies in boxes 1 and 2 if the overall number of blue galaxies were declining with time.  Number density evolution can occur through mergers of systems or by migration of blue galaxies to the red sequence.  It is interesting that the luminous end of the radius-luminosity relation shown in Figure \ref{fig:RL} tends to be populated by redder galaxies at all redshifts.  Results from recent work on the evolution of the luminosity function for luminous blue galaxies (Faber et al. 2005) indicate that such a number density evolution, if happening, is small.  Therefore, we will assume constant number density for this paper.

More complex evolutionary scenarios are possible.  For instance, the shape of the R-L distribution at a given size may change.  This would be the result if a subset of galaxies of a given size experienced a luminosity enhancement at high redshift, while the bulk of galaxies of that size remain in a 'normal' state.  However, our relatively small sample precludes exploration of these more complicated scenarios. 

The rest of this section will be dedicated to quantifying the evolution indicated in Figure \ref{fig:RL} out to $z=1$.  The figure indicates that the evolutionary trends continue beyond $z=1$, however, because of small number statistics, we choose to limit ourselves to galaxies below this redshift limit. We use the  previously defined SDSS sample of low-$z$ galaxies for a local reference. When comparing SDSS to TKRS, we seek to avoid the confusion that results from working with  large unequal volume bins.   For instance, when examining the case for radius or luminosity evolution, the use of equal volume bins allows us to monitor changes in number density.  

\subsection{Luminosity Evolution}
\begin{figure}
\includegraphics[trim=0 0 0 0, scale=0.65]{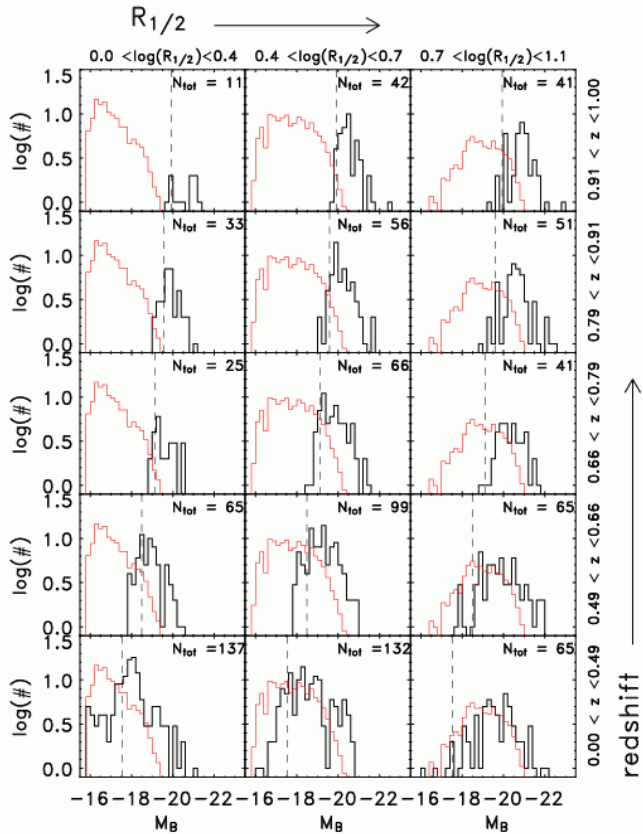}
\figcaption{\label{fig:lum_only} TKRS number distributions of $M_B$ (black) separated into 3 size groups and 5 equal volume ($\sim3 \times 10^4$ Mpc$^3$) bins out to $z=1$.  The red curves are drawn from the local SDSS sample and are plotted in each $z$-bin for comparison. The SDSS distributions are scaled to the same co-moving volume as TKRS. The TKRS distributions are shifted toward higher luminosities relative to SDSS at increasing redshift.  The TKRS magnitude limit of the upper end of each redshift bin is indicated by a vertical dashed line.
}
\end{figure}

First we investigate the case of luminosity evolution. Figure \ref{fig:lum_only} replots the TKRS R-L distributions (black), as histograms of galaxy magnitude, binned by size, and equal volume redshift bins to $z=1$.   The the local SDSS distributions (red) are separated into the same size and magnitude ranges.  The amplitude of the SDSS distributions are scaled to match the same volume as the TKRS distributions ($\sim3 \times 10^4$ Mpc$^3$) and are plotted in each $z$-bin as a reference.  These plots indicate that, in the lowest-$z$ bin, the SDSS distributions are similar to TKRS with some modest luminosity evolution.  As redshift increases, the peaks of the TKRS distributions shift towards higher luminosities.

Several effects should be noted while examining these distributions.  The sharp cut-off on the faint end of the GOODS distributions (especially at higher-$z$) is the result of the apparent magnitude limit of the TKRS survey.  The SDSS data, however, are complete to $M_B = -16$, meaning that the observed turnover of the SDSS distributions at fainter magnitudes for large- and medium-sized galaxies is real.  Although this turn over is not revealed for the small systems, the steep decline in numbers of objects at the bright end is an alternative marker for tracking evolution.  

Assuming that the SDSS sample represents the local universe, what is the likelihood that the TKRS sample was drawn from the local distribution?  Figure \ref{fig:lum_only} indicates that the two samples differ significantly.  If, however, we were to develop models for luminosity evolution and apply them to the SDSS curves, we could test which models produce the best match between the two distributions.   

\begin{figure}
\includegraphics[trim=80 150 0 350,clip,scale=0.5]{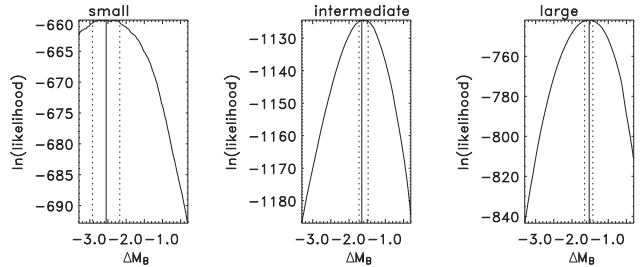}
\figcaption{\label{fig:maxlike}  ln(likelihood) vs. $\alpha=\Delta M_B(z=1)$ is plotted for the three galaxy size bins.  In order to minimize the effects of noise, the likelihood curves are boxcar smoothed with smoothing length of 0.1 magnitudes. The peak of each curve (solid vertical line) represents the most likely evolution, while the errors are given by a ln(likelihood) drop of 0.5 (dotted lines).  Given the input model, large and intermediate sized galaxies have evolved by $1.53 (-0.10, +0.13)$  and $1.65 (-0.18, +0.08)$ magnitudes respectively.  Small galaxies appear to have evolved by an additional magnitude, with $\Delta M_B(z=1) = 2.55 (\pm 0.38)$ mags, different from the results for larger galaxies at about the 2 sigma level. If we do not smooth the likelihood curves or we fit for the peak, the results are virtually identical and well within the errors.
}
\end{figure}

We have chosen a simple model where magnitude evolves linearly with redshift, such that 
\begin{equation}
\label{eqn:mshift}
\Delta M_B(z) = \alpha \cdot z. 
\end{equation}
This model assumes that the shapes of the magnitude distributions do not change with redshift;  it is a pure shift in luminosity for galaxies of a given size.  We use a maximum likelihood method to recover the most likely $\alpha$ for galaxies in each size bin.  We begin by  assuming a value for $\alpha$.  For each TKRS galaxy in a given size bin, we shift the SDSS number distribution by $\Delta M_B(z)$ to the TKRS galaxy redshift.  We then cut the SDSS number distribution at the TKRS magnitude limit for that redshift, and  normalize it to unity creating a probability distribution. We  calculate the likelihood of this TKRS galaxy being in our shifted SDSS sample.  Likelihood is given by the probability associated with a galaxy of given $M_B$ and $R_{1/2}$ read from the shifted and normalized SDSS probability distribution, multiplied by the galaxy's TKRS selection bias (top number from Figure \ref{fig:rl_ap}).  Finally, we sum ln(likelihood) for all galaxies at all redshifts in a given size bin. We  repeat this process with a range of values for $\alpha$ to find the most likely magnitude shift between the SDSS and TKRS distributions.  

Figure \ref{fig:maxlike} plots ln(likelihood) vs. $\alpha$ and reveals the luminosity evolution experienced by galaxies of each size bin to $z=1$. In order to decrease noise, the likelihood curves are boxcar smoothed with smoothing length of 0.1 magnitudes.  The most likely evolution is given by the peak of each curve (solid line), while the representative errors (dashed lines) are given by the locations where ln(likelihood) drops by 0.5 (for a description of the maximum likelihood method see Kendall, Stuart \& Ord 1987).  Interestingly, galaxies of different sizes seem to be  experiencing different amounts of luminosity evolution.  While the analysis indicates that large and intermediate-sized galaxies have experienced roughly equivalent evolutions of $1.53 (-0.10,+0.13)  $ and $1.65 (-0.18, +0.08)$ magnitudes respectively, small galaxies appear to have experienced significantly more, $2.55 (\pm0.38)$ mag.   The result for galaxies with small size is different from the results for larger galaxies at about the 2 sigma level, suggestive of a radius dependent luminosity evolution.  If we do not smooth the likelihood curves or we fit for the peak, the results are virtually identical and well within the errors.

\begin{figure}
\includegraphics[trim=100 130 0 160,scale=0.65]{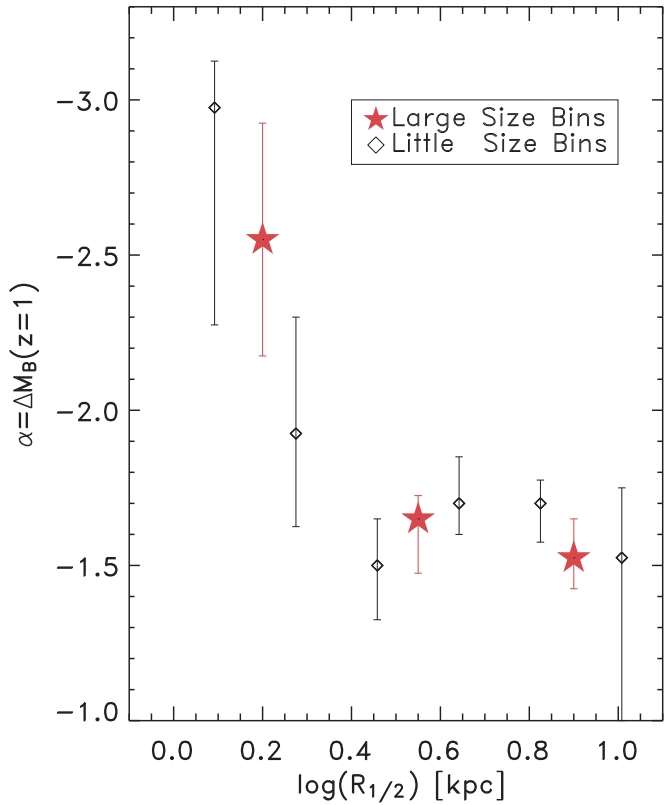}
\figcaption{\label{fig:rad_dep_lum}  A plot of the dependence of $\alpha$ on galaxy size.  We repeat the maximum likelihood method with smaller radius bins (black points) with width 0.18 in log$(R_{1/2})$.  The results for the larger, original radius bins are repeated in red.  A chi squared test of the black points rules out, at the 2 sigma level, an evolutionary model with constant luminosity evolution for galaxies of all radii. This suggests that the upturn in $\alpha$ at small radii is real.  Larger samples, especially of small galaxies, will help to test these result. 
}
\end{figure}

We check whether the change in $\alpha$ with radius is abrupt by repeating the maximum likelihood technique on smaller size bins.   The results, shown in Figure \ref{fig:rad_dep_lum}, indicate that a transition may be occurring for objects with sizes smaller $\sim2$ kpc.   Larger sample sizes, especially for small galaxies will help to confirm or disprove the claim of size dependence on the luminosity evolution.  This may be possible in on-going and future redshift surveys such as the Deep Extragalactic Evolutionary Probe 2 (DEEP2). 

\begin{figure}
\includegraphics[trim=0 0 0 0,scale=0.65]{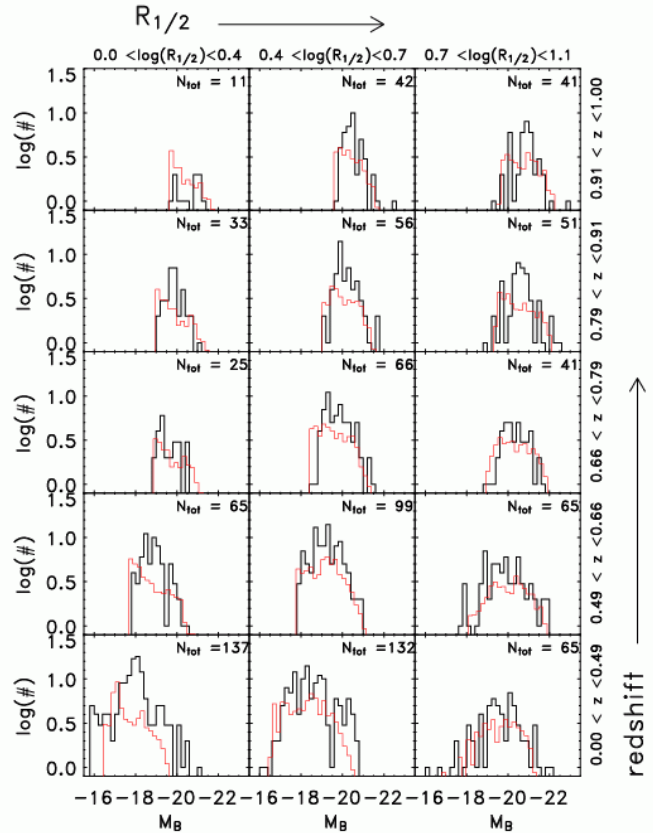}
\figcaption{\label{fig:lum_only_se} Similar to Figure \ref{fig:lum_only}, but now the SDSS distributions (red) are shifted in magnitude to match the median redshift of the bin, and the TKRS selection function is applied.  The match between the TKRS and evolved SDSS distributions provides a visual confirmation  of the maximum likelihood fit to luminosity evolution.
}
\end{figure}

Figure \ref{fig:lum_only_se} shows the same TKRS (black) distributions as in Figure \ref{fig:lum_only}, but now  the SDSS (red) distributions are shown with the simple luminosity evolution model and the TKRS selection function applied.  Thus the red curves now represent the distribution our model predicts for TKRS, given the evolved SDSS distribution.  Aside from small differences in the total number of objects, the red and black curves match well.  The differences in number density could reflect an actual evolution in number density that we are not accounting for in our models.  It could also result from overestimating the selection bias.  For example if a large sample of the galaxies in Figure \ref{fig:rl_ap} are at higher redshift than the galaxies targeted by TKRS, then we would be overestimating the selection bias.  Other factors might contribute to the differences in the two samples, for instance cosmic variance, or some more complex evolutionary scenario.  Deeper studies with larger sample sizes may help to identify the causes for these differences.

\begin{figure}
\includegraphics[trim=0 0 0 0,scale=0.65]{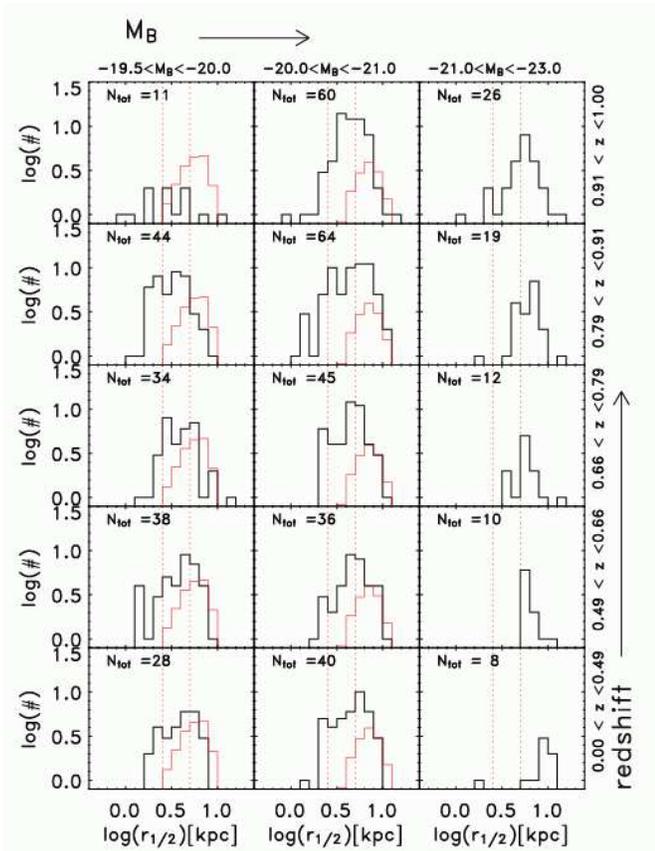}
\figcaption{\label{fig:rad_only} TKRS number distributions in half-light radius (black) for three different luminosity classes and equal volume bins to $z=1$.  The magnitude bins were chosen such that the low luminosity bin is complete out to $z=0.9$. The red vertical lines at $R_{1/2} = 2.5$ and 5 kpc are drawn for comparison.  There is no clear trend of increasing radius with time (at fixed magnitude) among galaxies in these luminosity classes.  However, there is a decline in the number density, as illustrated by the lack of SDSS galaxies (red) in the most luminous bin, consistent with luminosity evolution rather than size evolution.  
}
\end{figure}  

Some of the measured evolution may be the result of changing active galactic nuclei (AGN) populations within the TKRS sample.  If we remove all Chandra sources (Chandra Deep Field North; Alexander et al. 2003)  from TKRS (89 galaxies within $1.5 \arcsec$ of a Chandra source), there is a modest change in the results.  The measured evolutions for large, intermediate and small galaxies is then $1.33 (\pm0.13,)$; $1.48(-0.10,+0.13)$; and $2.23(-0.33,+0.18)$ respectively.  This indicates that most of the evolution is not the result of a change in the number density of AGN. 

\subsection{Radius Evolution}

We now perform a similar analysis to look for evolution in the half-light radius, dividing the sample into 3 luminosity bins, Figure \ref{fig:rad_only}.   Once again the TKRS data are shown in black and the SDSS data are shown in red.  Because the redshift bins are equal volume, we can readily see that this plot is not consistent with pure radius evolution.  If pure radius evolution were occurring, the total number of galaxies within a given luminosity bin would be constant over time, while  the peaks of the distributions would shift to larger radii.  Instead, the total numbers of objects within each magnitude bin are declining over time, with the largest decline in the brightest magnitude bin.  In fact, in the most luminous bin, SDSS predicts no galaxies within the given volume at the present day.  In contrast, TKRS shows that there are 24 galaxies in this same luminosity bin by $z=1$.  Note, the highest redshift, lowest luminosity bin in Figure \ref{fig:rad_only} is compromised by the magnitude limit of TKRS.

\begin{figure}
\includegraphics[trim=0 0 0 0, scale=0.65]{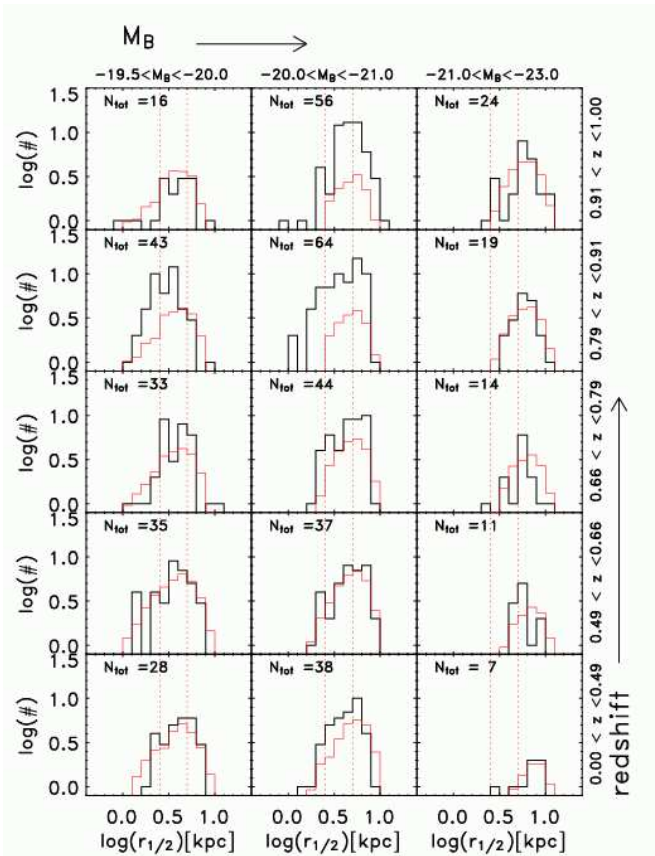}
\figcaption{\label{fig:rad_only_ev}  We re-plot Figure \ref{fig:rad_only} only now we apply our measured luminosity evolution and the TKRS selection functions to the SDSS sample (red) and, except for 2 bins, find it is a good match to TKRS(black).   }
 \end{figure}

Figure \ref{fig:rad_only} is consistent with significant luminosity evolution.  The drop in number of luminous objects with time is a prediction of luminosity evolution.   The fact that the SDSS and TKRS samples do not match well in the low-$z$ bin is also explainable by the significant luminosity evolution predicted in the previous section.  Recall that by $z=0.5$ we are predicting more than half a magnitude of luminosity evolution.  Figure \ref{fig:rad_only_ev} reproduces Figure \ref{fig:rad_only} now applying our measured radius dependent luminosity evolution and TKRS selection function to the SDSS sample.   Again pure  luminosity evolution is shown to be a good explanation for the differences between the SDSS and TKRS samples. Note two intermediate luminosity bins ($ 0.79< z < 1.0$) show fewer SDSS galaxies than expected.  This may be another indication for more complicated evolutionary scenarios.  

\section{Discussion}
We have demonstrated evolution in the radius-luminosity relationship for blue galaxies.  Based on number density constraints, this evolution cannot be explained as a pure radius evolution.  It can, however, be explained by a radius dependent luminosity evolution.  We have modeled this evolution as a simple shift in luminosity that increases linearly with redshift.  In this section we describe how our results compare to previous work and attempt to provide a physical mechanism that might give rise to the measured luminosity evolution.  We first examine the results from large and intermediate sized galaxies which appear to be experiencing similar evolution, roughly a factor of 4 decrease in luminosity since $z=1$.  We then discuss the more surprising result from galaxies with small size that appear to have evolved by nearly a factor of 10 in luminosity.

\subsection{Evolution of Large and Intermediate Sized Galaxies} 
A consistent picture is beginning to emerge for the evolution of large and intermediate sized blue galaxies.   Our results for large and intermediate sized galaxies are well matched to the findings in Barden et al. (2005), who examined the R-L relation in the GEMS field.  They found a central $B-$band surface brightness evolution of $\sim 1.4$ magnitudes for disk galaxies.  Our measured evolution is consistent though slightly larger, perhaps due to different selection criteria.  We selected all blue galaxies, while Barden et al. selected galaxies with an exponential Sersic profile, so we may be including more centrally concentrated objects such as AGN.  Our sample includes 89 x-ray sources, which may be harboring central AGN.  If we remove these sources from our sample we measure a somewhat smaller evolution ($\sim0.2$ mags for each size bin), in better agreement with Barden et al.  It should be noted, however, that the differences between our results and Barden et al. are small compared with the uncertainties.  



Our results are also well matched to Trujillo \& Aguerri (2004) who looked for surface brightness evolution disk galaxies within the HDF.  Assuming that the surface brightness evolution is driven by luminosity evolution they find a $V$-band luminosity evolution of roughly 0.8 magnitudes by z=0.7.   Assuming some small color evolution of $0.1-0.2$ mags, suggested by Figure \ref{fig:RL}, this is, to within errors, what we predict for large and intermediate sized galaxies ($\sim1.1 \pm 0.2$ mags by $z=0.7$).  Trujillo \& Aguerri (2004) also investigated a size-only evolution model.    They used the SDSS galaxy sample of Shen et al. (2003) as a local baseline for the R-L relationship, but they did not consider the number density of galaxies in given size and luminosity bins.  Specifically, because Shen et al. encompasses a much larger volume than the HDF, the samples especially for bright systems are not comparable.  As we show when volumes are considered, the radius-only evolution interpretation does not work. 

In addition to luminosity evolution Barden et al. (2005) also showed evidence for constant stellar mass surface density in disk galaxies since $z=1$.  They interpreted this as evidence for inside out disk growth.  If we adopt a 25\% growth in half-light radius for our galaxies since $z=1$ and apply it to our SDSS reference sample, we can remeasure the expected amount of additional  luminosity evolution in the sample.  As a simple model, we assume that the size evolution is linear with redshift and that all galaxies are growing at the same rate.  The resulting luminosity evolutions since $z=1$ are then $1.18 (-0.05,+0.13)$, $1.33 (\pm 0.10)$, and $1.75 (-0.18, +0.45)$ for large, intermediate, and small galaxies respectively.  Under this scenario, a significant amount of the surface brightness evolution is from the increasing galaxy size, and therefore the measured luminosity evolution is smaller.  Thus, if galaxies really have grown in size with time, our measurements of luminosity evolution from the previous section are upper limits.

Trujillo \& Pohlen (2005) looked for evidence of size evolution in disk galaxies.  They measured luminosity and truncation radii in 21 of 36 disk galaxies in the UDF.   They next applied the 1.4 magnitude luminosity evolution suggested by Barden et al. to their high-$z$ R-L relation.  They found it was still offset from their own measured local R-L relation (their Figure 2).   The offset in the two samples was largely the result of the four smallest galaxies in their high-$z$ sample.  The offset could be from truncation radius growth, or it could be from additional luminosity fading.  They claim that an additional 25\% growth in truncation radius can account for the offset.  If, however, these smaller galaxies were evolved by our measured luminosity evolution (2.55 mags since $z=1$ for small galaxies),  they would lie on the local relation without invoking size evolution.  It is not clear how truncation radius and half-light size are related, and our sample is significantly different from theirs, so it is not clear if our luminosity evolution is appropriate for these systems. It is, however, reassuring to see a similar behavior in both samples (Trujillo \& Pohlen and ours), small galaxies require more evolution than large ones.  

Further investigations of the Barden et al. (2005) claim of inside-out disk growth should be made.  
For instance, it would be helpful to directly measure the spatial distribution of star formation in intermediate redshift galaxies.  Where are the dominant sites of star formation?  Are they predominantly on the outskirts of galaxies with disks growing from the inside out? Are they smoothly distributed throughout the disks?  Are they centrally concentrated possibly driven by minor mergers and interactions? As galaxies become more bulge dominated they might drop out of the Barden et al. (2005) sample because of higher Sersic indexes.  This might bias the Barden et al. measurement.  We can learn something about the spatial distributions of star formation from the ACS $B$-band images which probe rest-frame UV.  However, because much of the star formation can be dust obscured especially by $z=1$, such studies might be biased against detecting centrally concentrated star formation.  A better way of detecting the sites of ongoing  star formation would be $H \alpha$ imaging with a combination of \HST\ ACS and NICMOS narrow-band filters.  $H\alpha$ will be less obscured by dust than UV, allowing us to identify the major sites of star formation.  Such a study would help us to understand how much inside out disk growth has occurred since $z=1$.  Unfortunately such a data set does not currently exist.

\subsection{Models of Luminosity Evolution}

Accepting that significant luminosity evolution has occurred within the blue galaxy population, can we construct a simple, consistent model that describes the evolving stellar populations of these galaxies?  For instance, is it possible for a stellar population to evolve by over 1.5 magnitudes in luminosity without evolving in color to the red sequence? Figure \ref{fig:model_ev} shows the $B$-band magnitude and $B-V$ color evolution of 3 simple stellar population models with exponentially declining star formation rates (SFR), or 'tau models', where tau gives the $e$-folding time-scale for decline in SFR.  The red model is effectively an instantaneous burst.  The green curve is a tau model of exponentially declining star formation with $\tau=3$ Gyr. The blue curve is effectively a constant star formation model.  
A time span of 8 Gyr, roughly the time between $z=1$ and today is marked off above the plot. Note that burst start time strongly effects the evolution measured for the instantaneous burst , but the extended evolution of the $\tau=3$ model is less dependent on start time. Over an 8 Gyr time span the $\tau=3$ Gyr (green) model declines in luminosity by $\sim1.5$ magnitudes.  In addition the $\tau=3 $ model remains on the blue sequence.  Thus a simple exponential decline in star formation (not necessarily a unique solution) can explain the decline in luminosity of intermediate and large sized galaxies in our sample, while keeping them blue.  

Based on the decline in optical luminosity, the models predict a drop by a factor of $\sim10$ in the SFR for the large and intermediate sized galaxies.   This result ties well to recent observations in the mid-infrared (MIR) with the Spitzer MIPS instrument.  The MIR has been shown to be well correlated with dusty star formation (e.g. Kennicutt 1998). Le Floch et al. (2004) among others have used the Spitzer data to identify a rapidly evolving population of luminous infrared galaxies (LIRGs) out to $z=1$.  While relatively common at $z=1$, LIRGs are rare today.   Melbourne, Koo \& Le Floc'h (2005) showed that the majority of LIRGs are blue galaxies and at high redshift the LIRGs are dominated by large normal-looking spiral morphology. Locally, LIRG spirals are very rare.  Based on changes in MIR luminosity, Figure 2 of Melbourne, Koo \& Le Floc'h also shows a qualitative decline in the SFR for large disks, on the order of the $\tau=3$ model.

\subsection{Evolution of Small Galaxies}
\begin{figure}
\includegraphics[trim=0 0 0 0,scale=0.85]{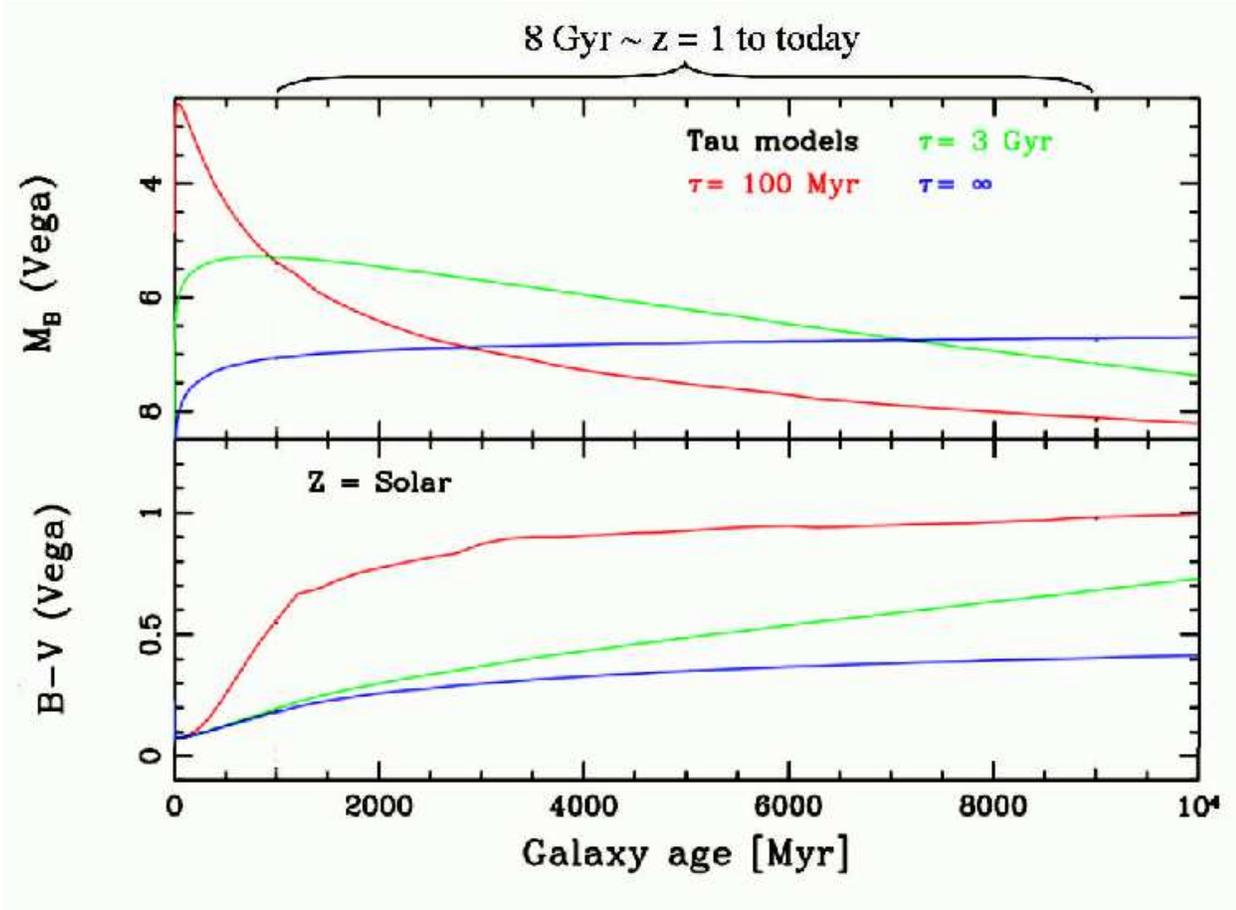}
\figcaption{\label{fig:model_ev} The luminosity (top) and color (bottom) evolution of solar metallicity stellar population synthesis models from Pegase (Fioc et al. 1997).  The red model is effectively an instantaneous burst.  The green curve is a tau model of exponentially declining star formation with $\tau=3$ Gyr. The blue curve is effectively a constant star formation model.  A time scale of 8 Gyr, roughly the amount of time between $z=1$ and today, is marked off above the plot.  Note that burst start time strongly effects the evolution measured for the instantaneous burst , but the extended evolution of the $\tau=3$ model is less dependent on start time.  Over the 8 Gyr time scale, the $\tau=3$ model declines by $\sim1.6$ mags and remains blue.  This model is a reasonable explanation for the luminosity evolution of large and intermediate sized TKRS galaxies. A shorter tau is necessary to describe the evolution of galaxies with smaller radii which appear to have evolved by $\sim2.5$ magnitudes since $z=1$.   
}
\end{figure}

The picture for small blue galaxies is not as simple as for the galaxies with larger size.  Our luminosity evolution model suggests that small blue galaxies have faded by a factor of 10 in luminosity since $z=1$.  This is a large amount of fading. Figure \ref{fig:model_ev} shows that in order to drop the luminosity of a galaxy by 2.5 magnitudes with tau models requires a very short tau, resulting in a huge decline in SFR over the last 8 Gyr.   For instance, a $\tau=1$ Gyr model results in SFR decline of over 1000 times since $z=1$, a very different regime from the results for larger galaxies.  Because small blue galaxies today tend to be sites of significant star formation, on the order of $0.3 M_{\sun} \; yr^{-1}$ for blue compact dwarfs (Hopkins et al. 2002), there is a problem fading the entire population of small blue galaxies by 2.5 magnitudes.  Local blue compact dwarfs would have to have been LIRGs and ultra-LIRGs at $z=1$, inconsistent with Melbourne, Koo and Le Floc'h (2005). Thus a simple tau model is not sufficient to explain the evolution of small blue galaxies as a group.  
  
For small systems we need to invoke more complicated evolutionary scenarios than pure luminosity evolution.  For instance, it is possible that the luminosity function shapes for small galaxies are not constant with redshift.  We may be seeing a subset of galaxies experiencing a significant star burst rather than a brightening of the entire ensemble of small galaxies at high $z$.  The only way to determine the behavior the entire ensemble of small galaxies is to study a sample that reaches fainter luminosities.  This may be possible in the ultra-deep field, but is beyond the scope of this paper.

The conclusion from the stellar population models in Figure \ref{fig:model_ev} is that the assumption of non-evolving shape for the SDSS luminosity distributions for small galaxies is probably wrong.  If we allow the shape to change then only a sub-sample of galaxies are experiencing large luminosity enhancements at high $z$.   The bursting dwarf systems seen at high redshift may fade significantly and are probably not the same objects that are bursting at lower redshift.  Interestingly the small galaxies that happen to be bursting at high redshift achieve higher luminosities than the bursting galaxies at lower redshift.   So some sort of evolution is taking place, either the intensity of the bursts or the frequency of the bursts has changed.  The details of this evolution, however, are not well constrained by our data. 

While the evolution inferred for the small galaxies is large, it is not a total surprise.  Over a decade ago, Koo et al. (1995) described a class of luminous compact blue galaxy (LCBG) at intermediate redshifts, with characteristics similar to our small radius sample. Essentially all of the small TKRS galaxies with  $z > 0.66$ are LCBGs.  A study of the LCBGs in HDF-N and the flanking fields by Phillips et al. (1997) found that  these galaxies make up as much as 10\% of the luminous galaxies seen at redshift 1.  However they are very rare today (Werk et al. 2004). Guzman et al. (1997)  inferred that these galaxies may account for as much as 40\% of the SFR evolution between $z=1$ and today.  Based on narrow line widths in spatially resolved STIS spectra (Bershady et al. 2005), and stellar mass estimates of $5\times10^9 M_{\sun}$  from infrared colors (Guzman et al. 2003), Bershady et al. concludes that many of these galaxies are low mass and may fade by as much as 4 magnitudes by today. In that context, a luminosity evolution  of over two magnitudes in the TKRS sample is easier to understand.  LCBGs are thought to be undergoing an unsustainable star burst, and are \emph{not} believed to represent the bulk of dwarf systems at high-$z$, but rather those that have experienced significant luminosity enhancement.  The LCBG scenario favors a changing shape to the luminosity distribution of small galaxies, rather than a simple shift in luminosity.  Again the proper way to fully explore what is happening is to obtain a sample that reaches to significantly lower luminosities, which may be possible in the UDF.  

\section{Conclusions}

We measured the size and luminosity of 1440 galaxies from the Team Keck Redshift Survey of GOODS-N.  Based on an an analysis of the UDF and same region analyzed at the GOODS depth we found that the GOODS field was not missing significant numbers of low-surface brightness galaxies to the TKRS apparent luminosity limit.  This indicates that previous studies of incompleteness in the deep \HST\ survey fields, that were based on idealized model galaxies, may have over estimated the incompleteness.  Blue galaxies at intermediate redshifts, $z\sim1$, were easier to detect than smooth models because they usually contained sub-structure such as star forming regions.
   
Our study of the blue galaxy population of TKRS revealed that the radius-luminosity relation has evolved since $z=1$.  The observed evolution is inconsistent with a pure radius evolution model, but it can be explained by a radius dependent luminosity evolution.  Assuming a linear decline in $B$-band magnitude, galaxies with large radii ($R_{1/2} > 5$ kpc) have evolved by $\Delta M_B(z=1)=1.53 (-0.10,+0.13)$.  Intermediate sized galaxies have experienced similar though slightly more evolution  of  $\Delta M_B=1.65 (-0.18, +0.08)$.  These declines in luminosity can be explained by a simple exponential decline in star formation.  In particular an exponentially declining SFR with $\tau=3$ Gyr is a good match to the fading and SFR decline that is measured. Since pure luminosity evolution can explain the observations while a pure radius evolution can not, it is reasonable to assume that luminosity evolution dominates.  This does not, however, preclude some moderate size growth, in which case we have measured upper limits to the luminosity evolution.

Small galaxies ( $R_{1/2} < 2.5$ kpc) appear to have undergone significantly more evolution, $\Delta M_B=2.55 (\pm0.38)$, almost a factor of 10 in luminosity, and different from larger galaxies at the two sigma level.  This sharp decline in luminosity cannot be explained by a simple exponential decline in star formation in the entire small galaxy population.  Tau models would predict a decline of roughly 1000 in SFR if the decline in luminosity were happening uniformly to the entire sample over the last 8 Gyrs.  Because small blue galaxies are sites of significant star formation today, this scenario does not work.   The luminosity decline can, however, be explained by a sub-population of bursting dwarf systems at high redshift that have faded by the current epoch.  Such a population has been previously identified as luminous, compact, blue galaxies that are rare locally but were relatively common by $z=1$.  The measured evolution could then be a result of a change in the intensity or frequency of bursts. 

\acknowledgements
We would like to thank Luc Simard for providing the local SDSS sample; Jay Strader and Chien Peng for key insightful comments; the DEEP team for long discussions of the project; and Jason Prochaska and Yasmin Lucero for help with statistical tests. 

This work was supported in part by the NSF Science and Technology Center for Adaptive Optics managed by UC Santa Cruz under the cooperative agreement No. AST-9876783. Partial funding came from the DEEP2 program supported by NSF grants AST-0071198, AST-0522950, and AST- 0071048.  Partial funding also came from the HST Archival grant HST-AR-009946.

This work used observations from the NASA/ESA Hubble Space Telescope, obtained 
from the data archive at the Space Telescope Science Institute. STScI is operated by the 
Association of Universities for Research in Astronomy, Inc. under NASA contract NAS 5- 
26555.  Redshifts came from the DEIMOS spectrograph funded by a grant from CARA (Keck Observatory), an NSF Facilities and Infrastructure grant (AST92-2540), the Center for Particle Astrophysics and by gifts from Sun Microsystems and the Quantum Corporation. These data were obtained at the W.M. Keck Observatory, which is operated as a scientific partnership among the California Institute of Technology, the University of California and the National Aeronautics and Space Administration. The Observatory was made possible by the generous financial support of the W.M. Keck Foundation. We also wish to recognize and acknowledge the highly significant cultural role and reverence that the summit of Mauna Kea has always had within the indigenous Hawaiian community.

\clearpage
\LongTables
\begin{deluxetable}{lccccccccccc}
\tabletypesize{\footnotesize}
\tablewidth{0pt}
\tablecaption{Elliptical Apperture Photometry of TKRS Galaxies \label{table:ktrs}}
\tablehead{
\colhead{                 TKRS\_ID\tablenotemark{1}} & \colhead{               R.A.(2000)\tablenotemark{1}} & \colhead{                     Dec.\tablenotemark{1}} & \colhead{                        z\tablenotemark{1}} & \colhead{               $m_B$ [AB]} & \colhead{                    $m_V$} & \colhead{                    $m_i$} & \colhead{                    $m_z$} & \colhead{             $M_B$ [Vega]} & \colhead{           $(B-V)_{rest}$} & \colhead{        $r_i$ [$\arcsec$]} & \colhead{      $R_B$ [kpc]} }
\startdata
    117&     12:36:55.897&      62:21:38.17&     0.585&     - &    23.781&    23.213&    23.040&   -18.642&     0.397&     0.668&     4.467\\
    239&     12:36:20.286&      62:17:29.60&     0.848&     - &    23.724&    22.678&    22.120&   -20.421&     0.648&     0.381&     2.848\\
    320&     12:36:32.725&      62:18:44.77&     0.483&     - &    24.274&    23.656&    23.845&   -17.660&     0.411&     0.220&     1.074\\
    345&     12:35:43.068&      62:12:55.89&     0.851&    24.552&    23.663&    22.859&    22.008&   -20.402&     0.733&     0.540&     4.238\\
    428&     12:37:05.851&      62:22:30.74&     0.487&    24.357&    23.792&    23.201&    23.696&   -18.154&     0.412&     0.536&     2.804\\
    432&     12:36:36.116&      62:19:05.01&     0.681&    23.716&    22.693&    21.759&    21.450&   -20.507&     0.561&     0.632&     4.551\\
    444&     12:36:01.540&      62:15:13.41&     0.298&     - &    22.085&    21.487&    21.183&   -18.416&     0.581&     0.918&     3.863\\
    445&     12:36:51.450&      62:20:53.53&     0.744&    22.923&    22.886&    21.984&    21.729&   -20.591&     0.407&     0.943&     6.739\\
\enddata
\tablenotetext{1}{Wirth et al. 2004}
\end{deluxetable}
\clearpage




\end{document}